\documentclass[intlimits,twoside,a4paper]{article}
\usepackage{amsmath,amssymb}
\usepackage{graphicx}
\usepackage{wrapfig}

\usepackage[T2A]{fontenc}
\usepackage[cp1251]{inputenc}
\usepackage[eqsecnum]{cmpj}

\issue{2011}{14}{4}{43802}

\doinumber{10.5488/CMP.14.43802}

\title[Graphene growth]%
{Growth of graphene on 6H--SiC \\by molecular dynamics simulation%
}
\author[N. Jakse, R. Arifin, S.K. Lai]{N. Jakse\refaddr{label1},
        R. Arifin\refaddr{label2}, S.K. Lai\refaddr{label2}}
\addresses{
\addr{label1} Sciences et Ing\'enierie des Mat\'eriaux et
Proc\'ed\'es, Grenoble INP, UJF-CNRS, \\ 1130 rue de la Piscine,
BP 75, 38402 Saint-Martin d'H\'eres Cedex, France \addr{label2}
Complex Liquids Laboratory, Department of Physics, National
Central University, Chungli 320, Taiwan }

\date{Received October 29, 2011, in final form November 20, 2011}

\authorcopyright{N. Jakse, R. Arifin, S.K. Lai, 2011}

\begin{document}

\maketitle

\begin{abstract}
Classical molecular-dynamics simulations were carried out to study
epitaxial growth of graphene on \linebreak 6H--SiC(0001) substrate. It was
found that there exists a threshold annealing temperature above
which we observe formation of graphitic structure on the
substrate. To check the sensitivity of the simulation results, we
tested two empirical potentials and evaluated their reliability by
the calculated characteristics of graphene, its carbon-carbon
bond-length, pair correlation function, and binding energy.
\pacs 81.05.ue 31.15.xv 68.55.A-
\end{abstract}

\section{Introduction}
Though many experiments were made on various properties of
graphene~\cite{Per09}, limited works have been reported on
applying the computer simulation technique to study the growth of
graphene. One very recent work of this kind was done by Tang et
al.~\cite{Tan08} who used classical molecular dynamics (MD)
simulation to grow graphene epitaxially on 6H--SiC substrate. The
numerical procedure which was used for growing graphene on
substrate, i.e. by removing layers of Si atoms to mimic Si
sublimation, is somewhat different from those recent experimental
observations by Poon et al.~\cite{Pon10} who put forth a step-flow
growth from edges of terraces. Nevertheless, their simulations
give a beneficial access to single and multi-layer graphene
formation from duly prepared carbon-rich layers. Subsequently,
simulation studies by the same authors address the thermal
stability of graphene on 6H--SiC substrate as well~\cite{Tan08b},
while other simulation works~\cite{Var08,Lan10} focus more on the
issue of thermal stability of the buffer carbon layer. As far as
classical simulations are concerned, the reliability of the
simulations depends on the quality of the interaction models, and
it is therefore of primary importance to assess their capability
of treating this system.

This paper reports a simulation study addressing the following
important issues: (a) growing graphene epitaxially on the
6H--SiC(0001) substrate, (b) evaluating graphene sheets formed on
the SiC substrate via comparing two different empirical potentials
one of which, the Tersoff potential~\cite{Ter89}, is widely used
in the literature and another, its modified version~\cite{Erh05},
is supposedly more accurate, and (c) discussing the structural
stability of mono- and two-layer graphene, based on the
characteristic features such as the carbon-carbon (C--C) pair
correlation function, bond length and binding energy. Attempt is
made to relate the simulated results to the experimentally
observed data.

\section{Molecular dynamics method}

\subsection{Numerical procedure: LAMMPS software and empirical potentials}

We conduct the MD simulations using the Large-scale
Atomic/Molecular Massively Parallel Simulator (LAMMPS)
software~\cite{Lam95}. Two separate MD simulations are carried
out, one of which employs the Tersoff potential~\cite{Ter89} and
the other one uses the modified Tersoff potential~\cite{Erh05}
(hereafter referred to as TEA). It should be noted that while the
functional form of both  potentials is the same, their
parameterizations are different, and even their respective range
of interaction, that is controlled through a smooth cutoff function
with two parameters of the potential, are somewhat different. We
refer the readers to the original papers for more details.
Interestingly, the parameter files of both these potentials are
available in LAMMPS library~\cite{Lam95}. To proceed to
simulations, we first describe how our input data are prepared,
and the use of them to study the growth of graphene on
6H--SiC(0001) substrate.

\subsection{Numerical procedure: molecular dynamics simulation}

\begin{wrapfigure}{i}{0.48\textwidth}
\centerline{
\includegraphics[width=0.47\textwidth]{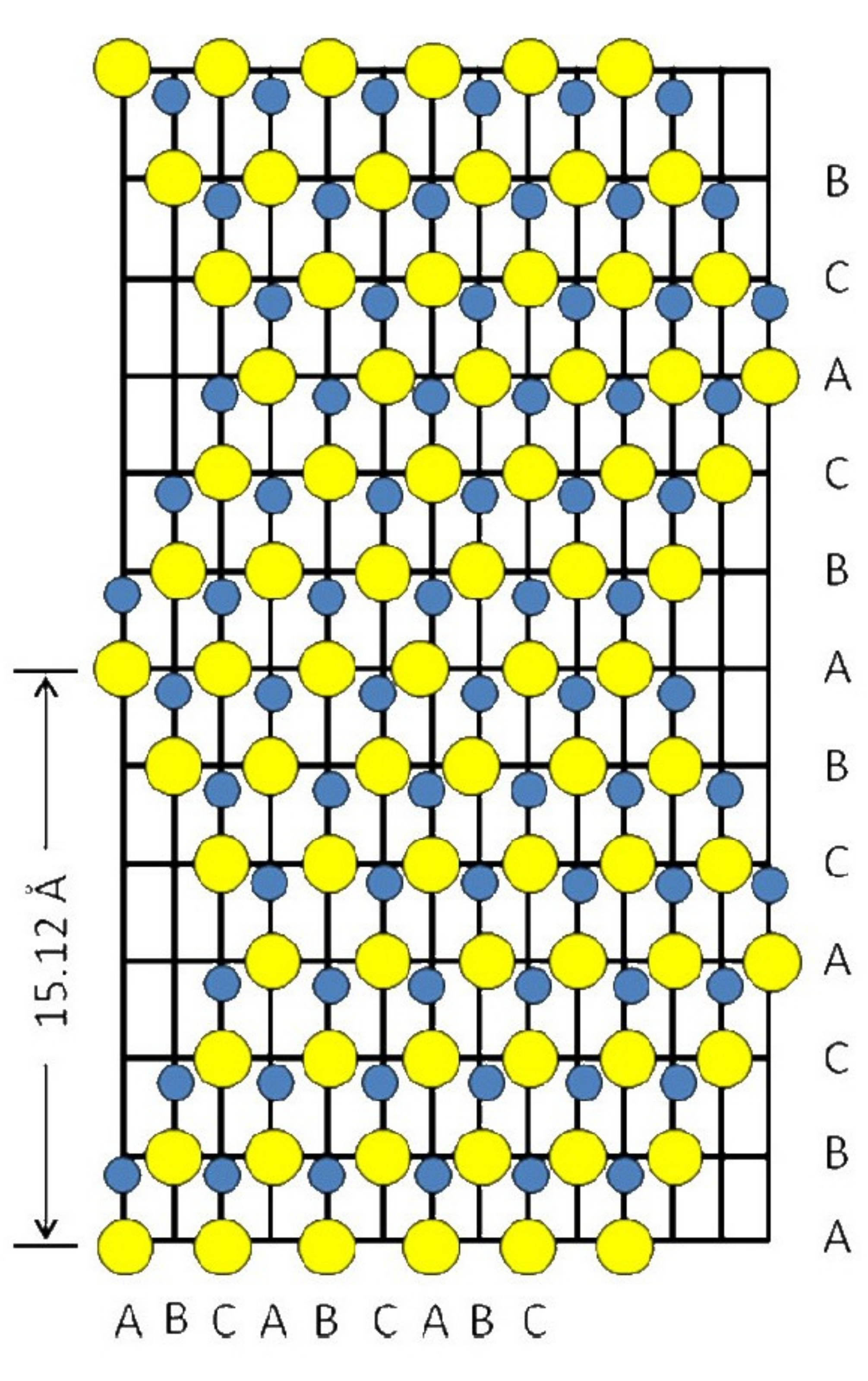}}
\caption{(Color online) Schematic diagram of Si-terminated
6H--SiC(0001). Small blue circles are carbon atoms while the large
yellow ones are Si atoms.} \label{figure1}
\end{wrapfigure}

In our Nose-Hoover (NVT ensemble) simulations of the graphene
growth, configurations of layers of carbon-rich atoms are
positioned to loll near a 6H--SiC substrate. Such configurations
are prepared in the following way. The SiC substrate is set using
a crystalline structure with six hexagonal layers repeating
periodically in the (0001) direction, each hexagonal layer
consisting of two sublayers, one for silicon and the other for
carbon. Thus, the stacking of 6H--SiC will thus run as ABCACB\ldots,
which is 6H--SiC(0001). In this work, we focus on the Si-terminated
6H--SiC (see figure~\ref{figure1})
%
%
and obtain C-rich layers by simply removing the topmost Si-layers,
which is a numerical procedure introduced to mimic the sublimation
process of Si atoms in the epitaxial growth of graphene. The
dimensions of our orthorhombic simulation cell containing the
6H--SiC substrate are $60.07\times61.36\times15.12$~\AA$^3$ in $x$
(generated with lattice parameter $2.668$~\AA), $y$ (generated
with lattice parameter $1.54$~\AA) and $z$ directions,
respectively. Note that the periodic boundary conditions are
applied along $x$ and $y$, while a vacuum of $30$~\AA \ is created
along the $z$ direction.

In addition to preparing the initial configurations of atoms, a
technical point which concerns the growth of multilayer graphene
is in order. In removing silicon atoms directly from the 6H--SiC
crystalline substrate, one focuses on the number of C-rich layers
by properly choosing prescribed distances among C-rich layers and
substrate. Consider, for example, four C-rich layers after
removing Si atoms from the 6H--SiC crystal. We keep the distance
between the substrate and the first-layer C-rich atoms next to it
at an original separation of $1.9$~\AA. Then, between the next
two, i.e. first and second C-rich layers, we make it to lie within
$1$~\AA \ or a separation smaller. We set a distance $1.9$~\AA \
between the second and third C-rich layers and resume a separation
of $1$~\AA \ again between the third and fourth C-rich layers.
Note that the C-rich layers take on the original crystalline
structure with Si atoms removed, i.e. a centered hexagonal
structure with a C--C bond-length of $2.78$~\AA \ which is longer
than $2.65$~\AA \ of Tang et al.~\cite{Tan08}. The stringent
condition between the C-rich layers and substrate should be
strictly obeyed otherwise only a few hexagonal rings are created.

\begin{figure}[!b]
\begin{center}
\begin{tabular}{cc}
\includegraphics[width=0.35\textwidth]{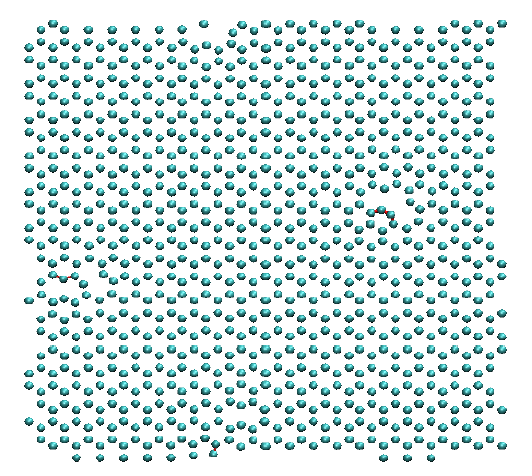} & \includegraphics[width=0.35\textwidth]{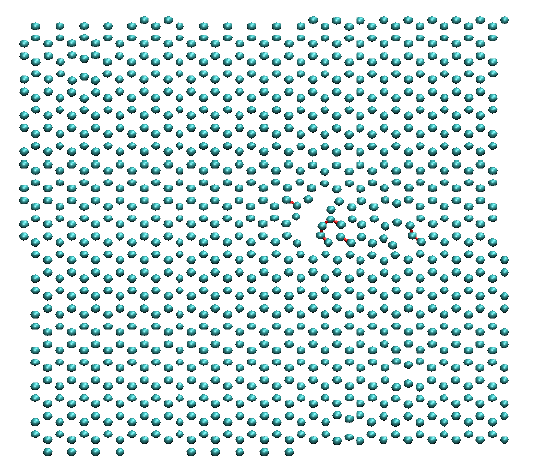} \\
1200 K & 1300 K \\
\includegraphics[width=0.37\textwidth]{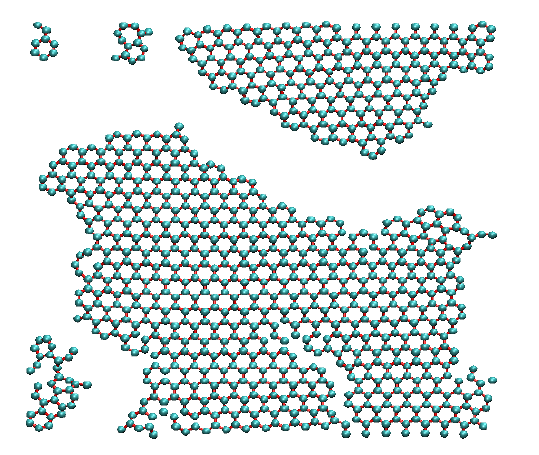} & \includegraphics[width=0.35\textwidth]{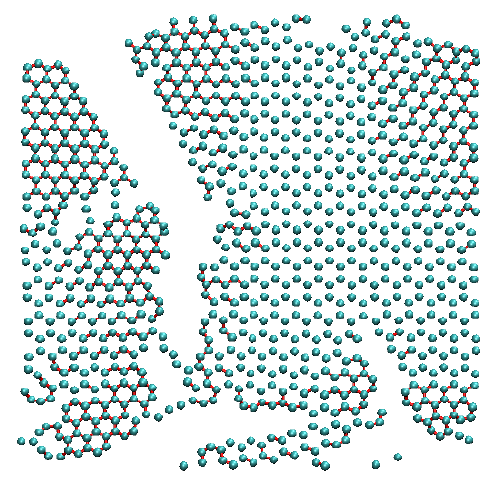} \\
1260 K & 1450 K
\end{tabular}
\end{center}
\caption{(Color online) Growth of monolayer graphene on 6H--SiC
substrate simulated with TEA (left) and Tersoff (right)
potentials. For TEA potential, the threshold annealing temperature
at which graphene emerges occurs at a lower temperature within the
range $1200<T_{\mathrm{tr}}<1260$~K, whereas it is higher within
the range $1300 < T_{\mathrm{tr}} < 1450$~K for Tersoff
potential.} \label{figure2}
\end{figure}

The procedure and the details of parameters used in this
simulation growth of graphene on 6H--SiC substrate are summarized
as follows:

\begin{itemize}

\item We apply the Tersoff and TEA potentials in two separate
simulation series to calculate the interatomic C--C and Si-C
interactions with parameters given in references~\cite{Ter89}
and~\cite{Erh05}, respectively.

\item We consider 6H--SiC as a substrate, i.e. six bilayers of Si
and C atoms and in each layer of Si or C atoms, a total number of
$480$ atoms is considered.

\item The simulation procedure is performed by first relaxing the
system using the conjugate gradient minimization
method~\cite{Pay92}. The initial distance of C--C atoms in the
C-rich layers after relaxation is $2.78$~\AA \ which is larger
than the $2.65$~\AA \ of Tang et al.~\cite{Tan08}. Then we perform
a MD simulation with a timestep  $\Delta t=0.5$~fs, and heat the
system until $T=300$~K is reached. In this process, the
temperature is increased using a linear ramp within the
Nose-Hoover thermostat, so that a heating rate of $10^{13}$~K/s is
imposed. At this temperature, the system is equilibrated for a
time interval of $2\times10^4\Delta t$. From this configuration, a
set of simulations are carried out to increase the temperature of
the system to various desired $T$, in order to study the
temperature evolution of the carbon layers. As above, a heating
rate of $10^{13}$~K/s is chosen. At each target temperature $T$,
equilibrium of the system is obtained after a total time interval
of $3\times10^4\Delta t$. In the final stage, the system is
annealed down to $T=0.1$~K at a cooling rate of $5\times10^{12}$~K/s
so that the properties of the carbon sheets can be studied without
thermal noise.
\end{itemize}

\section{Results and discussion}

\subsection{Growth of monolayer graphene on 6H--SiC substrate}

The monolayer graphene simulated with the TEA and Tersoff
potentials are compared in figure~\ref{figure2}. The main feature
to emphasize is the existence of a threshold annealing temperature
$T_{\rm tr}$ signaling the emergence of graphene. For the TEA potential,
the carbon layers start to transform at 1200~K and the graphene
layer is formed at $T_{\rm tr} = 1260$~K. The latter is close to the
experimentally value observed by Hannon and Tromp~\cite{Han08} who
report the formation of smooth steps of graphene in prolonged
annealing at 1298~K. For Tersoff's potential, the transformation
of the carbon layers to graphene sheet spreads from 1300~K to the
threshold value of 1450~K, which is a more gradual formation of
graphene and at a higher temperature than the
experiment~\cite{Han08}.
\begin{figure}[!htb]
\begin{center}
\begin{tabular}{cc}
\includegraphics[width=0.45\textwidth]{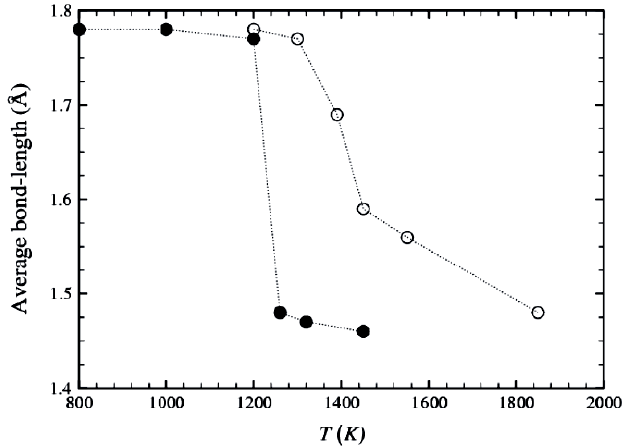} & \includegraphics[width=0.45\textwidth]{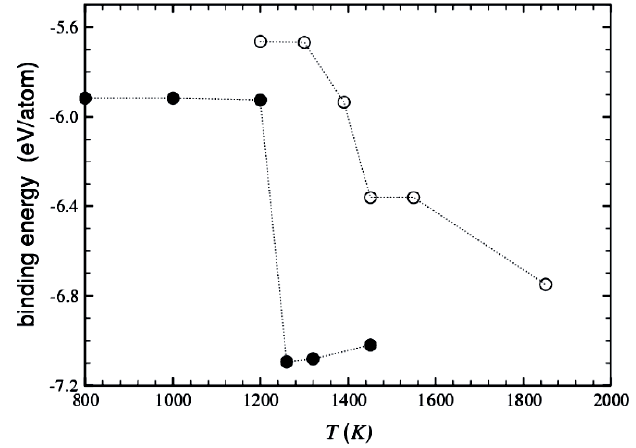} \\
\includegraphics[width=0.45\textwidth]{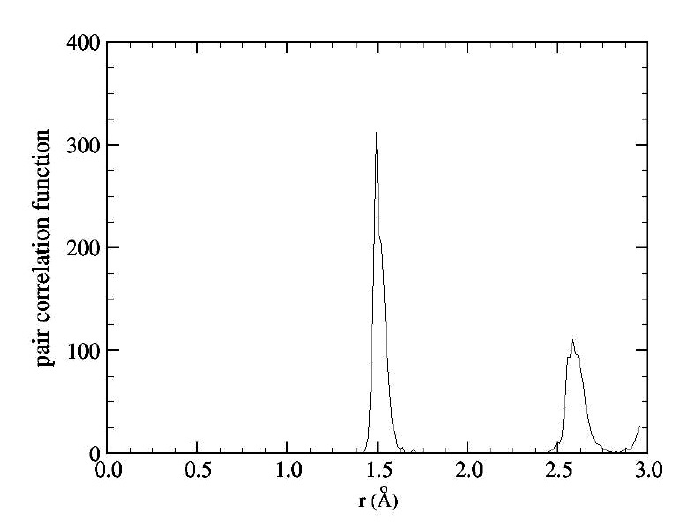} & \includegraphics[width=0.45\textwidth]{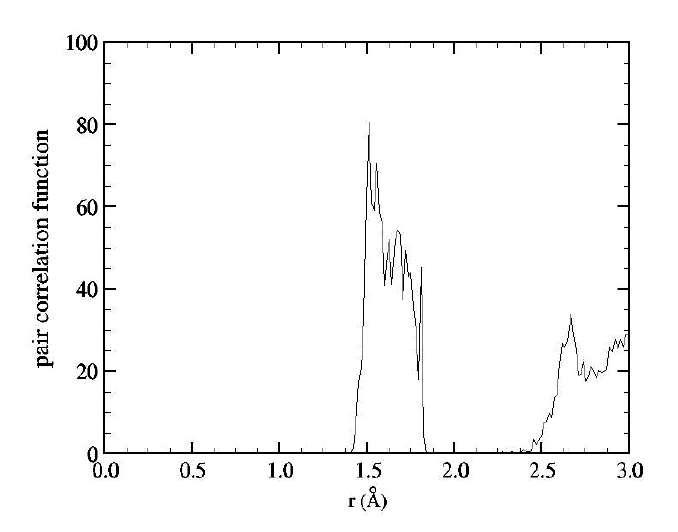} \\
\end{tabular}
\end{center}
\caption{Top row: Comparison of the average C--C bond-length (left)
and binding energy (right) of monolayer graphene simulated using
the Tersoff (open circle) and TEA (solid circle) potentials.
Bottom row: The pair correlation function of graphene obtained by
TEA potential (left) at $1260$~K and Tersoff potential (right) at
$1450$~K for a monolayer graphene grown on 6H--SiC
substrate.} \label{figure3}
\end{figure}

\subsection{Bond-length, binding energy, and pair correlation function of monolayer graphene}

The concrete evidence that the TEA potential yields a well-defined
graphene structure is its prediction of an average C--C bond-length
equal to $1.48$~\AA \ (at $1260$ K) (see figures~\ref{figure2}
and~\ref{figure3}). This value is close to the $sp^2$-hybridized
graphitic carbon ($1.42$~\AA). As for Tersoff's potential, a value
of $1.59$~\AA \ (at $1450$ K)~\cite{Com1} is obtained, which is
not as good as the TEA potential. The pair correlation functions
$g(r)$ in figure~\ref{figure3} are also consistent with the
results of bond-length; the position of the first maximum of
$g(r)$ is $1.487$~\AA \ for the TEA potential to be compared with
$1.508$~\AA \ for the Tersoff potential. Note in
figure~\ref{figure3} that the binding energy for the former
potential is $-7.0941$~eV/atom, which is lower than the
$-6.3620$~eV/atom obtained from Tersoff's potential. It is worth
mentioning that the binding energies have been calculated directly
from the potential energy per carbon atom given by the Tersoff or
TEA potential. Armed with these results, our study for two-layer
graphene grown on SiC substrate will proceed below by using only
the TEA potential.
\begin{figure}[!b]
\begin{center}
\begin{tabular}{ccc}
& {\bf 1000 K} & \\
First layer &  &  Second layer\\
\includegraphics[width=0.35\textwidth]{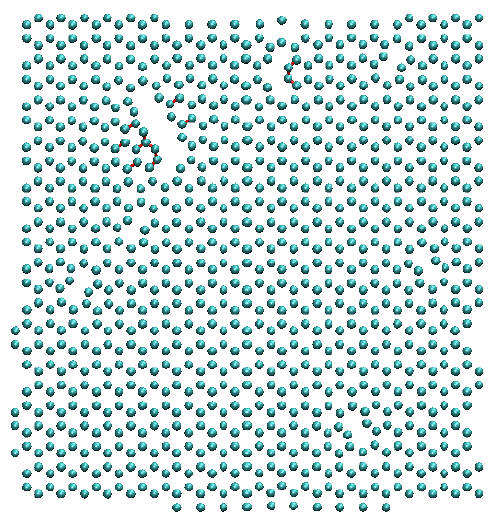} & & \includegraphics[width=0.35\textwidth]{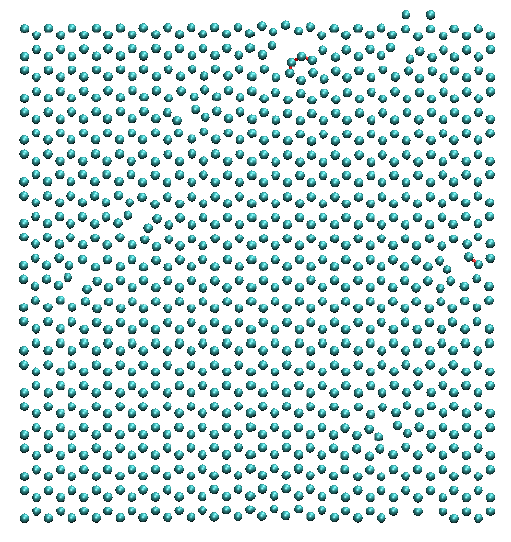} \\
& {\bf 1320 K} & \\
First layer &  &  Second layer\\
\includegraphics[width=0.35\textwidth]{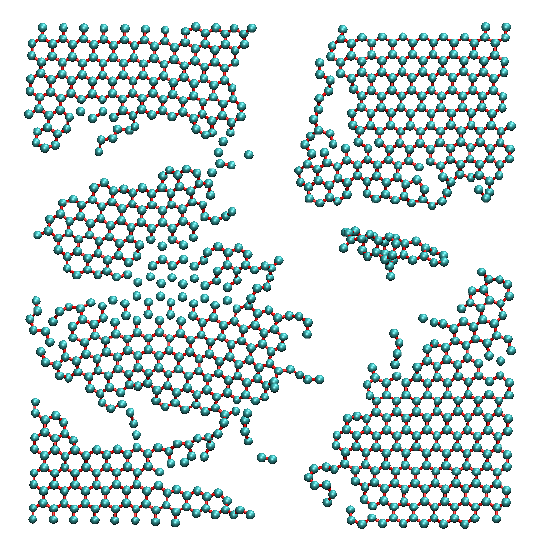} & &  \includegraphics[width=0.35\textwidth]{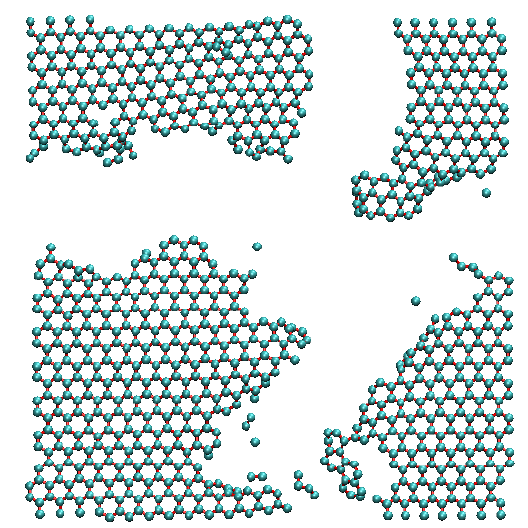}
\end{tabular}
\end{center}
\caption{(Color online) Two layers of graphene grown on 6H--SiC
substrate simulated with TEA potential. The first-layer graphene
corresponds to the one near the substrate.} \label{figure4}
\end{figure}

\begin{figure}[h]
\centerline{\includegraphics[width=0.49\textwidth]{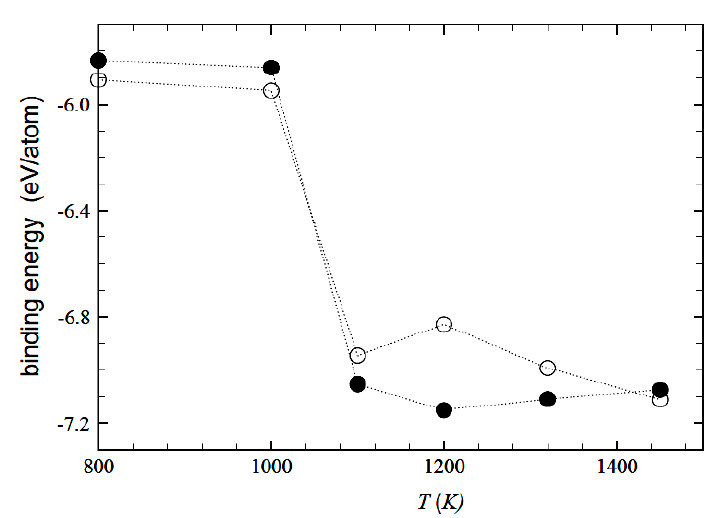}
\hfill
\includegraphics[width=0.47\textwidth]{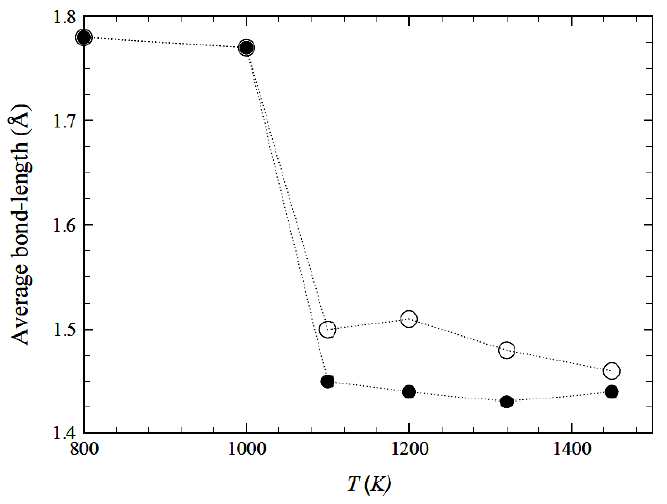}}
\caption{Average bond-length (right) and binding energy (left) for
the first (open circle) and second (solid circle) layers graphene
grown on 6H--SiC substrate.} \label{figure5}
\end{figure}

\subsection{Growth of two-layer graphene on 6H--SiC substrate}

Figure~\ref{figure4} shows the two-layer graphene grown on 6H--SiC
substrate. The first graphene layer, which is the one that clings
to substrate, is relatively less stable than the second layer.
This is evident from examining Figure~\ref{figure5} for the
bond-length and binding energy for the two layers. The bond-length
of the second graphene layer at $T=1320$ K is $1.43$~\AA. This
value is smaller than $1.48$~\AA \ of the first layer and is
closer in magnitude to the $1.42$~\AA \ bond-length of
$sp^2$-hybridized graphene carbon. Further evidence can be gleaned
also from the pair correlation function $g(r)$ at $T = 1320$ K
whose first maximum position for the first layer is $1.53$~\AA \
to be compared with $1.48$~\AA \ of the second layer. The binding
energy of the second layer at $T=1320$ K is estimated to be
$-7.1085$ eV/atom which is lower than $-6.9925$ eV/atom for the
first layer.

\section{Conclusion}

We have applied the MD simulation to the study of the growth of
graphene on 6H--SiC substrate by epitaxial method. It was found
that the choice of an empirical potential in simulation is
sensible in predicting the threshold temperature at which point
the graphene starts to emerge. With the TEA potential, we obtained
one and two layers of graphene grown on 6H--SiC(0001) substrate. In
addition to yielding the threshold annealing temperature, which
was found to be reasonably close to that implied in recent
experiments, the characteristics of the grown graphene are
confirmed by the calculated average C--C bond-length,
pair-correlation function and binding energy. With this empirical
potential in hand, one can envisage investigating the thermal
stability of multilayer graphene and carrying out a deeper
analysis of growth mechanisms. Works along these lines are in
progress.

\section*{Acknowledgements}

The author would like to thank the National Science Council of
Taiwan for financial support (NSC100--2119--M--008--023). We are
grateful to the National Center for High-performance \linebreak {Computing},
Taiwan for computer time and facilities. The supercomputing
resource center \linebreak Phynum/CIMENT is gratefully acknowledged for providing
computing time.

\ukrainianpart

\title{Ріст графену на 6H--SiC  в симуляціях методом \\ молекулярної
динаміки}

\author{Н. Жакс\refaddr{label1}, Р. Аріфен \refaddr{label2}, С.К. Лаі\refaddr{label2}}

\addresses{
\addr{label1}  Національний політехнічний інститут Ґренобля, Університет Жозефа Фур’є--Національний центр наукових досліджень, 38402 Сан-Мартен д’Ере, Франція
\addr{label2}  Лабораторія складних рідин, Фізичний факультет, Національний центральний університет, Чунґлі 320, Тайвань
}

\makeukrtitle

\begin{abstract}

Для того, щоб дослідити епітаксіальний ріст графену на підкладці
6H--SiC(0001) здійснено симуляції методом класичної молекулярної
динаміки. Знайдено існування  порогу температури відпалу, вище якої спостерігається формування на підкладці
структури графену. Для того, щоб перевірити чутливість результатів
симуляцій, ми тестуємо два емпіричних потенціали і оцінюємо їхню
надійність шляхом розрахунку характеристик графену, довжини зв'язку
вуглець-вуглець, парної кореляційної функції та енергії зв'язку.
\keywords графен, епітаксіальний ріст, молекулярна динаміка
\end{abstract}

\end{document}